\documentclass[aps,prd,nofootinbib,twocolumn,longbibliography]{revtex4-1}

\usepackage{amsmath,amssymb}
\usepackage{graphicx}
\usepackage[bookmarks=false,pdfstartview=FitH]{hyperref} 

\def\tr{{\rm tr}}
\def\x{{\mathbf x}}
\def\be{\begin{equation}}
\def\ee{\end{equation}}

\begin{document}

\title{\large General relativistic statistical mechanics}

\author{Carlo Rovelli}
\affiliation{Centre de Physique Th\'eorique de Luminy, Aix-Marseille University, F-13288 Marseille, EU}

\begin{abstract}

\noindent Understanding thermodynamics and statistical mechanics in the full general relativistic context is an open problem.  I give tentative definitions of equilibrium state, mean values, mean geometry, entropy and temperature, which reduce to the conventional ones in the non-relativistic limit, but remain valid for a general covariant theory.   The formalism extends to quantum theory.  The construction builds on the  idea of thermal time, on a notion of locality for this time, and on the distinction between global  and  local temperature. The last is the temperature measured by a local thermometer, and is given by $kT = \hbar\, {d\tau}/{ds}$, with $k$ the Boltzmann constant, $\hbar$ the Planck constant, $ds$ proper time and $d\tau$ the equilibrium thermal time. 

\end{abstract}

\maketitle

\section{Introduction}

Thermodynamics and statistical mechanics are powerful and vastly general tools.  But their usual formulation works only in the non-general-relativistic limit. Can they be extended to fully general relativistic systems? 

The problem can be posed in physical terms: we do not know the position of each molecule of a gas, or the value of the electromagnetic field at each point in a hot cavity, as these fluctuate thermally, but we can give a statistical description of their properties. For the same reason, we do not know the \emph{exact} value of the gravitational field, which is to say the exact form of the spacetime geometry around us, since nothing forbids it from fluctuating like any other field to which it is coupled. Is there a theoretical tool for describing these fluctuations?

The problem should not be confused with thermodynamics and statistical mechanics on curved spacetime. The difference is the same as the distinction between  the dynamics of matter on a given curved geometry versus the dynamics of geometry itself, or the dynamics of charged particles versus dynamics of the electromagnetic field. Thermodynamics on curved spacetime is well understood (see the classic  \cite{Tolman}) and statistical mechanics on curved spacetimes is an interesting domain (for a recent intriguing  perspective see \cite{Smerlak:2011yc}).  The problem is also distinct from ``stochastic gravity" \cite{Hu:1989db,Hu:2003qn}, where metric fluctuations are generated by a Einstein-Langevin equation and related to semiclassical effects of quantum theory.   Here, instead, the problem is the just the thermal behavior of conventional gravity.\footnote{One may ask whether  equilibrium can ever be reached, given the gravitational instabilities and  long thermalization times. The question is legitimate, but doesn't authorize us evading the issue of what equilibrium means: for the question itself to even make sense, and because we are always concerned only with approximate equilibrium in nature, gravity or not.}    

A number of puzzling relations between gravity and thermodynamics (or gravity, thermodynamics and quantum theory) have been extensively discussed in the literature \cite{Bekenstein:1974ax,Bekenstein:1973ur,Hawking2,Bardeen:1973gs,Wald:1994uq,Padmanabhan:2011zz,Padmanabhan:2003gd,Jacobson:2003vx,Jacobson:2003wv,Carlip:2012ff}. Among the most intriguing are probably Jacobson's celebrated derivation of the Einstein equations from the entropy-area relation \cite{Jacobson:1995ab,Jacobson:2012yt}, and Penrose Weil-curvature hypothesis \cite{Penrose:1979fk,Penrose:2006zz}.  These are very suggestive, but perhaps their significance cannot be evaluated until we better understand standard general covariant thermodynamics. 

One avenue for addressing the problem is perturbation theory. Another is restricting to asymptotic flatness and observables at infinity \cite{York:1986it,Brown:1992bq,Brown:1989fa}.  Although useful in specific contexts, these roads are incomplete, because they miss the core issue: understanding if temperature has a meaning in the bulk of spacetime in a strong field regime.  What do we mean when we say that near a cosmological singularity temperature is high? For the moment we do not have a definition of temperature that makes sense where the metric might fluctuate widely.

A step towards general covariant statistical mechanics was taken in \cite{Rovelli:1993ys,Rovelli:1993zz} and extended to quantum field theory in \cite{Connes:1994hv}.  The notion introduced in these papers is \emph{thermal time}. This is meant to address the basic difficulty of general relativistic statistical mechanics: in a generally covariant theory, dynamics is given relationally rather than in terms of evolution in physical time\footnote{For a  discussion of this crucial point see the Appendix and Chapter 3 of \cite{Rovelli:2004fk}, in particular Section 3.2.4.}, consequently the canonical hamiltonian vanishes, and without a hamiltonian $H$ it is difficult to even start doing statistical physics.  The idea of thermal time is to reinterpret the relation between Gibbs states ($\rho\propto e^{-\beta H}$) and time flow (generated by $H$): instead of viewing the Gibbs states as determined by the time flow, observe that any generic state generates its own time flow. The time with respect to which a covariant state is in equilibrium can therefore be read out from the state itself. The root of the temporal structure is thus coded in the non commutativity of the Poisson or quantum algebra \cite{albook,Connes:1994hv}.

Since any state is stationary with respect to its own flow, the problem left open is characterizing the states that are in \emph{physical} equilibrium.  Here we consider a solution: equilibrium states are those whose thermal time is a flow in spacetime.\footnote{This problem is considered also in \cite{Martinetti:2002sz,Longo:2009mn,Buchholz:1998pv}.} These, we suggest, are the proper generalization of Gibbs states to the general covariant context. 

This step allows temperature to be defined, following the intuition in \cite{Martinetti:2002sz,Rovelli:2010mv}: the temperature measured by a local clock is the ratio between thermal time and proper time. This yields immediately the Tolman-Ehrenfest law \cite{Tolman:1930zz, TolmanEhrenfest}, which correctly governs equilibrium temperature in gravity.  Entropy and free energy can be defined and we obtain the full basis of generally covariant thermodynamics. The construction extends to the quantum theory. 

The result is a tentative set of equations that generalize conventional thermodynamics and statistical mechanics to classical and quantum general-covariant systems.

\vspace{.2cm}
\centerline{---}
\vspace{.2cm}

We use units where the Boltzmann constant $k$ and the Planck constant $\hbar$ are set to unity. We have tried to keep the main text brief, confining background material to a detailed Appendix. The reader is urged to start from the Appendix unless the language and the background ideas of the text are already familiar. Equations in the paper are to be understood locally in phase space, namely on a chart  where suitable regularity conditions are satisfied to avoid singular or degenerate behavior. A finer analysis will make sense after the basic conceptual structure is clear.

\section{General covariant Gibbs states}

\subsection{Thermal time}

Let $\cal E$ be a symplectic space, whose physical interpretation is the extended phase space of a general covariant theory (see the Appendix for notation and details.)  Let $\cal C$ be a submanifold of  $\cal E$, representing the surface where the constraints  of the theory (which code the full dynamics) are satisfied.  The symplectic form $\sigma$ of $\cal E$ induces a presymplectic structure on $\cal C$, whose null directions can be integrated to define the gauge orbits $o$.  The space $\Gamma$ of these gauge orbits, which is the physical phase space of the theory,  is again a symplectic space, with symplectic form $\omega$.  It is in 1-to-1 correspondence with the space of the solutions of the field equations, modulo gauges.  A statistical state $\rho$ is a positive function on $\Gamma$ normalized with the respect to the Liouville measure 
\be
        \int_\Gamma \rho = 1. 
        \label{norm}
\ee 
The hamiltonian vector field $X$ defined by 
\be
 \rho \ \omega(X) = d\rho
    \label{X}
\ee
generates a flow $\alpha_\tau$ in $\Gamma$ called the thermal flow; its generator 
\be
    h=-\ln \rho
    \label{equi}
\ee
is called the thermal hamiltonian and the flow parameter $\tau$ is called thermal time\footnote{So defined, $\tau$ has the dimensions of an action, as it is conjugate to a dimensionless quantity. It can be made dimensionless by multiplying the r.h.s.\,of \eqref{X} and \eqref{equi} by $\hbar$. This is a bit artificial in the classical theory, but will be natural in the quantum theory.}  \cite{Rovelli:1993ys}. 

\subsection{Local thermal time}

Consider a general covariant theory that includes general relativity\footnote{We systematically disregard at this stage the difficulty of defining the Liouville measure that defines the integral \eqref{norm} in the case of field a theory. This is because the issue should properly be addressed in the quantum context, where I will be a bit more precise.}, and assume physical 3d space $\Sigma$ to be compact, with the $S_3$ topology. The space $\cal E$ can be coordinatized by the 3d Riemann metric tensor $q$ of $\Sigma$, the matter fields $\varphi$, and their respective conjugate momenta $(p,\pi)$; these quantities are fields on $\Sigma$, namely functions from $\Sigma$ to a target space $(q,\varphi,p,\pi):\Sigma\to V$.  An orbit $o$ determines a solution of the field equations and therefore in particular a pseudo-riemannian manifold $(M,g)_{\!o}$. A point in $o$ determines a spacelike Cauchy surface $\phi:S_3\to (M,g)_{\!o}$, having the given induced metric $q$ and extrinsic curvature $p$. In particular, a foliation $\phi_\tau:S_3\to (M,g)_{\!o}, \tau\in R$ of $(M,g)_{\!o}$ corresponds to a line on the orbit. 

Consider now a real function $\tilde T$ on $V$. This determines a local function (which we indicate with the same letter) on ${\cal E}$, namely a map $\tilde T:{\cal E}\times\Sigma\to R$ given by $\tilde T((q,p,\varphi,\pi),\x )=\tilde T(q(\x ),p(\x ),\varphi(\x ),\pi(\x )), \x \in\Sigma$.  The coordinate $\tilde T(\x)$ on $\cal E$ plays a role of ``multi-fingered time" in what follows. If the equation 
\be
\tilde T(\x )=\tau,  \hspace{6em} \tau\in R
\ee
defines a foliation of $(M,g)_{\!o}$ (on a given region of phase-space) we say that $\tilde T(x)$ is a ``local time".  The parameter of the foliation defines then a time coordinate $\tau:(M,g)_{\!o}\to R$ on spacetime.  The simplest example is if the matter fields include a scalar field $\tilde T$ that grows monotonically in spacetime (for the given region of phase space): then the value of the field defines a time coordinate. 

If there are canonical coordinates $\tilde T$ and $Q^i$ on ${\cal E}$, with respective momenta $P_{\tilde T}$ and $P_i$, such that 
\be
P_{\tilde T}(\x )=-h(Q^i(\x ),P_i(\x ))
\label{form}
\ee
on $\cal C$, then $\tilde T(\x )$ defines a deparametrization of the theory in the following sense: the hamiltonian 
\be
h=\int d^3\x \ h(Q^i(\x ),P_i(\x ))
\label{h}
\ee
evolves geometry and matter fields along the foliation $\phi_\tau$. Notice that $h$ is constant along the orbits it generates. We can therefore associate its value to each $o$ and obtain in this manner a function $h$ on $\Gamma$.  (Weaker cases are also of interest, in particular the case relevant in cosmology where 
\be
P_{\tilde T}(\x )=-f(\tilde T(\x ))\ h(Q^i(\x ),P_i(\x )); 
\label{form2}
\ee
which describe a system with temperature varying in time: see  \cite{Rovelli:1993zz}.)

Let us now come to the first main notion that we introduce in this paper. We say that a statistical state $\rho$ on $\Gamma$ is a ``Gibbs state" if there is a local time $\tilde T(x)$ with a local hamiltonian $h$ of the form \eqref{h} (or \eqref{form2}) satisfying \eqref{equi} up to an additive constant\footnote{The constant has no effect on the dynamics and we set it to zero by redefining $P_{\tilde T}$.}. 

If this is the case, the thermal time $\tau$ generated by $\rho$ is precisely the foliation time $\tau$, and therefore thermal time has a geometrical interpretation as a flow in spacetime.

\subsection{Nonrelativistic limit}

The definition above is  a generalization of the conventional definition of Gibbs states.  To see this, recall that for a Hamiltonian system with phase space $\Gamma_0$, canonical coordinates $(q,p)$ and Hamiltonian $H=H(q,p)$, a Gibbs state is a state of the form $\rho_\beta=Z^{-1}(\beta)\, e^{-\beta H}$ with $Z(\beta)\equiv\int_{\Gamma_0}e^{-\beta H}$. The general covariant formulation of this system is defined on the extended phase space ${\cal E}$ with canonical coordinates $(t,p_t,q,p)$ and the constraint $C=p_t+H(q,p)$. The constraint surface is coordinatized by $(t,q,p)$ and the orbits are given by $(t,q(t),p(t))$ where $q(t)$ and $p(t)$ are the solutions of the Hamilton equations. The space $\Gamma$ of these orbits is isomorphic to $\Gamma_0$ (but not canonically so, until a $t=t_0$ is chosen) via $(q,p)=q(t_0),p(t_0)$. 

A time function on ${\cal E}$ is provided by $\tau=t/\beta$, whose conjugate momentum is $p_\tau=\beta p_t$, which satisfies the requirement that the constraint can be expressed in the form \eqref{form}, namely $p_\tau=-h(q,p)$, where $h=\beta H$. The hamiltonian, being constant on each orbit, is well defined on $\Gamma$, therefore $\rho_\beta$ is a function on $\Gamma$, namely it is a statistical state in the covariant sense. It is immediate to see that it satisfies \eqref{equi}. In other words, the Gibbs state picks out the coordinate $t$ from $\cal E$, where this was confounded with the other variables. 

Observe now that the temperature $T\equiv\frac1{\beta}$ is equal to the ratio 
\be
T=\frac\tau{t}
\label{ttt}
\ee 
between the thermal time $\tau$, namely the parameter of the evolution generated by the logarithm of the Gibbs state, and the physical time $t$. This characterization of temperature can be extended to general covariant systems.

\subsection{Mean values, mean geometry\\ and local temperature}

Consider a family $\cal A$ of functions $A$ on $\Gamma$.  Let the mean value of $A$ on the state $\rho$ be
\be
\bar A = \int_\Gamma A\,\rho.
\ee
The thermal time flow $\alpha_\tau$ acts on these functions by $A(\tau)(s)=\alpha_\tau(A)(s)=A(\alpha_{-\tau}(s)), s\in \Gamma$, which satisfies  $dA/d\tau=\{A,h\}$.
Since $\rho$ is clearly invariant under the flow, so are the mean values, but 
\be
f_{AB}(\tau) = \int_\Gamma A(\tau)B\,\rho.
\ee
is in general a non trivial function and describes temporal correlations in the state.  Define the \emph{mean geometry} $\bar g$  (if it exists) of a state $\rho$ for an observable family $\cal A$ as a spacetime $(M,\bar g)$ with a foliation $\phi_\tau$ such that 
\be
      \bar A(\tau)=A(\phi^{-1}_\tau(\bar g)). 
\ee
Since $\bar A(\tau)$ is $\tau$ independent, it follows that $(M,\bar g)$ is stationary under the flow defined by $\phi_\tau$. Therefore $\xi=\frac{\partial}{\partial\tau}$ is a timelike Killing field on $(M,\bar g)$. The norm of $\xi$ is $ds/d\tau$ namely the ratio between the local flow of proper time and thermal time. The equivalence principle therefore compels us to define the local temperature by the local version of \eqref{ttt}, namely 
\be
\hspace{6em}  T(x) = |\xi(x)|^{-1}, \hspace{4em} x\in M
\ee
from which the Tolman-Ehrenfest law \cite{Tolman:1930zz,TolmanEhrenfest}
\be
   T(x)|\xi(x)|=constant
   \label{tolman}
\ee
that governs the spacetime variation of temperature at equilibrium in gravity, follows immediately.\footnote{A suggestion in this direction was in \cite{Martinetti:2002sz}. The intriguing relation between \eqref{ttt} and the Tolman law was pointed out in \cite{Rovelli:2010mv}.}  In stationary coordinates $(\tau, \x )$, the temperature is the inverse of the Lapse function, since $ds^2=N^2 d\tau^2$. 

\subsection{Partition function \\ and global temperature}

If $\rho$ is a Gibbs state, we can obtain another Gibbs state by exponentiating it with a constant $\beta$ and multiplying it by a $\beta$ dependent factor that preserves the normalization: $\rho_\beta=Z^{-1}(\beta)\, \rho^\beta$. The effect of this exponentiation is to scale the thermal time globally, and therefore to scale the temperature globally.  Therefore the global temperature is defined with respect to a reference Gibbs state. Having  a one-parameter family of Gibbs states allows us to define the partition function
\be
      Z(\beta) =  \int_\Gamma \rho^\beta. 
\ee
The entropy of the state can be obtained as usual from  
\be
      S(\beta) = - \int_\Gamma \rho_\beta \ln\rho_\beta 
\ee
and from this we can derive in a few steps the standard thermodynamical relation
\begin{eqnarray}
      S &=&- \int_\Gamma \rho_\beta (\beta \ln \rho-\ln Z)\nonumber \\
      &=&\beta E+\ln Z
\end{eqnarray}
where $E$ is the mean value of the energy $h\!=\!-\ln\rho$. The global temperature $\beta$ of the state should not be confused with the local temperature $T(x)$, which is space-dependent. Also, the local temperature is defined directly by a single statistical state (if a mean geometry exists), while the global temperature is only defined relative to another Gibbs state taken as reference. 

In the following section, we extend this structure to the quantum theory. 

\subsection{Quantum theory}

Let $\cal K$ be the unconstrained Hilbert space of a general covariant theory and $\cal H$ its physical Hilbert space (the ``space of solutions of the Wheeler-deWitt equation"). General covariant quantum mechanics is well defined by these structures (See the Appendix, and, more in detail, Section 5.2 of \cite{Rovelli:2004fk}.) 

A quantum statistical state is a trace-class operator $\rho$ on $\cal H$ such that $\tr\rho=1$. Its entropy is $S= -\tr[\rho\ln\rho]$.  Let $\cal A$ be an observable algebra formed by self-adjoint operators $A$ on $\cal H$. Then $\rho$ defines a state on this algebra by
\be
\rho(A)=\tr[A\rho]
\ee
and\footnote{Taking $\cal A$ to be a von Neumann algebra, namely  a ${}^*$-algebra of bounded operators closed in the weak operator topology and including the identity.} the Tomita theorem provides a flow $\alpha_\tau:{\cal A}\to{\cal A}$ on the observable algebra. This is the thermal-time flow in the quantum theory \cite{Connes:1994hv}. If there is a \emph{local} hamiltonian $h$ and a (now dimensionless) conjugate ``time" observable $\tau$ that  in the classical theory reduces to the quantities defined in the previous section, and generates an evolution 
\be
\alpha_\tau(A)=e^{\frac{i}{\hbar} h \tau}A\, e^{-\frac{i}{\hbar} h \tau}, 
\ee
then we say that $\rho$ is a Gibbs state.\footnote{Since space is compact, the usual difficulty of hamiltonian quantum  field theory with thermal states which historically gave rise to algebraic quantum field theory, is not there, since energy does not diverge on thermal states.}  The Tomita flow of $\rho$ satisfies the KMS condition (see,  for instance, \cite{Haag:1992hx})   
\be
  f_{AB}(\tau)=f_{BA}(-\tau+2\pi i)    \label{KMS}
\ee
for any two observables $A$ and $B$, where 
\be
  f_{AB}(\tau)=  \rho(\alpha_\tau(A)B). 
\ee
A thermal state $\rho_\beta=Z^{-1}(\beta)\,\rho^{\beta/2\pi}$ satisfies the KMS condition 
\be
  f_{AB}(\tau)=f_{BA}(-t+i\beta)    \label{KMS2}
\ee
with respect to the flow generated by $\rho$. 

The notion of mean geometry can be extended to the quantum theory\footnote{The idea of mean geometry is implicit in contexts where covariant quantum states of gravity are associated to a classical geometry \cite{Ashtekar:1992tm,Iwasaki:1992qy,Sahlmann:2001nv,Bianchi:2009ky,Conrady:2008ea,Livine:2006it}.} 
by defining $(M,\bar g, \phi_\tau)$  (if it exists) as the mean geometry of the state $\rho$ with respect to a given observable algebra $\cal A$ if
\be
\bar A(\tau)\equiv \rho(\alpha_\tau(A))) =A(\phi^{-1}_\tau(\bar g)).
\ee 
The local temperature $T(x)$ is defined by the norm of the killing field of the mean geometry, and is therefore a semiclassical concept. Restoring physical units, local temperature is given on the mean geometry by   
\be
 T(x) =  \frac{\hbar}{k}\  \frac{d\tau }{ds} , 
 \label{temperature}
\ee
where $\hbar$ is the Planck constant and $k$ is the Boltzmann constant. 

Notice that \eqref{temperature} gives the Unruh temperature \cite{Unruh:1976db} of a quantum field theory on Minkowski space, if $ds$ is the proper time along the accelerated observer trajectory and $\tau$ is the dimensionless parameter of the Bisognano-Wichman flow  $U(\tau)=e^{i\tau K/2\pi}$, where $K$ is the boost generator, which is the Tomita flow of the vacuum state on the Rindler-wedge observables \cite{Bisognano:1975fk,Haag:1992hx}.  

This suggests that the Unruh effect should affect the local temperature of an observer accelerated on a mean geometry, also in the context of the full generally-covariant statistical mechanics of the gravitational field.  If a mean geometry has a Killing horizon, where the norm of $\xi$ becomes singular, then the local temperature \eqref{temperature} diverges on the horizon. The divergence of the temperature is a high-energy, namely a short-distance phenomenon, therefore we can consider it in a region of spacetime small with respect to the local curvature of the mean geometry, namely as a locally flat-space phenomenon. As such, it must be determined by the Unruh temperature. An explicit example of a statistical state where this happens has been discussed in \cite{Bianchi:2012ui,BianchiStoccolma}. 
 An Unruh temperature in the vicinity of the horizon of a black hole is  red-shifted by the Tolman relation \eqref{tolman} precisely to Hawking's black hole temperature at infinity.

\section{Conclusion}

We have extended the machinery of statistical thermodynamics to the general covariant context. The new concepts with respect to conventional statistical mechanics are:
\begin{enumerate} 
\item The statistical state is defined on the space of the solution of the field equation.
\item Each statistical state defines a preferred time flow, called thermal time.
\item A statistical state whose thermal time flow has a geometrical interpretation, in the sense that it can be reinterpreted as evolution with respect to a local internal time, defines a generalized Gibbs state, with properties similar to the conventional equilibrium states.
\item For such states, it is possible to define the relative global temperature between two states. 
\item A mean geometry is a stationary classical geometry with a timelike killing field and a time foliation, such that the value of a suitable family of observables reproduces the statistical expectation values of these observables in the statistical ensemble.  
\item  If a mean geometry exists, a local temperature is defined. Local temperature is the ratio between proper time and thermal time on the mean geometry:
\be
 T(x) =  \frac{\hbar}{k}\  \frac{d\tau }{ds} , 
 \label{temperature2}
\ee
It yields immediately the Tolman law. 
\end{enumerate} 
This construction reduces to conventional thermodynamics for conventional Hamiltonian systems rewritten in a parametrized language. 

Examples, extension of the formalism to the boundary formalism \cite{Oeckl:2003vu,Rovelli:2002ef,Rovelli:2003ev}, which  is the natural language for quantum field theory in the generally covariant context, and applications to horizon thermodynamics, and in particular to the local framework defined in \cite{Frodden:2011eb} and the derivation of black hole entropy in loop quantum gravity in \cite{Bianchi:2012ui}, will be considered elsewhere. 

\centerline{---}

I thank Alejandro Perez for the crucial suggestion to focus on the locality of the hamiltonian, Ed Wilson-Ewing for pointing out the relevance for cosmology of the weaker notion of equilibrium, captured by \eqref{form2}, Eugenio Bianchi for numerous discussions on this subject,  Simone Speziale and Pierre Martinetti for several helpful comments. 

\vspace{1cm}

\section*{Appendix}

\subsection{Classical theory}

\subsubsection{Mechanics}

A conventional hamiltonian system is defined by a $2N$ dimensional phase space $\Gamma_0$ and a hamiltonian $H$. The phase space is a symplectic space, namely a manifold equipped by a non-singular closed symplectic two form $\omega$. Locally, we can always choose coordinates $(q^i,p_i)$ on $\Gamma$ such that 
\be
\omega=dq^i\wedge dp_i
\ee
(summation understood).  Having a symplectic two form is the same as having Poisson brackets. $H$ is a scalar function on $\Gamma$. Every function $f$ on a symplectic space defines a vector field $X_f$ on the space, defined by 
\be
\omega(X_f)=-df, 
\ee
where the l.h.s is the action of a differential two-form on a vector, which gives a one form, and the r.h.s is the differential of $f$. In turn, a vector field defines a flow $\alpha_t:\Gamma_0\to\Gamma_0, t\in R$, namely a continuous one-parameter group of automorphisms of $\Gamma_0$ into itself, related to $X$ by 
\be
\left.\frac{d\alpha_t}{dt}\right|_{t=0}=X_f. 
\ee
The Poisson bracket between two functions $A$ and $B$ on $\Gamma_0$ is defined by 
\be
\{A,B\}= X_B(A)=-X_A(B).
\ee
The flow of the hamiltonian is the time flow, namely the evolution in time of each point of $\Gamma_0$. Explicitly, the hamiltonian vector field of $H$ is easily seen to be
\be
X=\frac{\partial H}{\partial p_i} \frac{\partial}{\partial q^i} -\frac{\partial H}{\partial q^i} \frac{\partial}{\partial p_i},
\ee
so that the time flow is determined by the Hamilton equations 
\be
\frac{dq^i(t)}{dt}=\frac{\partial H}{\partial p_i} , \hspace{2em} \frac{dp_i(t)}{dt}=-\frac{\partial H}{\partial q^i}.
\ee
which show that this geometric construction is equivalent to hamiltonian mechanics.  An observable $A$ is a real function on $\Gamma_0$. The time evolution of an observable is defined by $A(t)=A\circ \alpha_t$ and satisfies 
\be
\frac{dA(t)}{dt}=\{A,H\}.
\ee

Let $\Gamma$ be the space of the solutions of the equation of motion $(q^i(t),p_i(t))$. This is a finite dimensional space with is isomorphic to $\Gamma_0$, but not canonically isomorphic. A specific isomorphism is obtained by choosing a value $t_0$ for the time parameter $t$. Then the isomorphism between  $\Gamma$ and  $\Gamma_0$ is given by $(q^i,p_i)=(q^i(t_0),p_i(t_0))$. Thanks to this isomorphism, $\Gamma$ has a symplectic structure as well (independent from $t_0$).            

For instance, the solutions of the dynamics of a harmonic oscillator have the form $(q(t)=A\sin(\omega t+\phi), \ \ p(t)=m\omega A\cos(\omega t+\phi))$. Therefore $\Gamma$ is coordinatized by $A$ and $\phi$. A map between $\Gamma$ and $\Gamma_0$ is obtaining choosing for instance $t=0$, which gives $(q=A\sin\phi,\ \  p=m\omega A\cos\phi)$ and therefore the symplectic form on $\Gamma$ is 
\be
\sigma=-m\omega\ A\, dA\wedge d\phi.
\ee

An equivalent formulation of the dynamics, called the presymplectic formulation, can be given on the $2N+1$ space ${\cal C}=\Gamma_0\times R$, with local coordinates $(q^i,p_i, t)$ equipped with the two-form
\be
\omega'=dq^i\wedge dp_i-dH(q^i,p_i)\wedge dt.
\ee
This two-form is degenerate, namely has a null direction (since the space is odd dimensional). That is, there exist a vector field $X'$, determined up to scaling, such that 
\be
\omega'(X')=0.
\ee
It is immediate to see that this vector field is proportional to
\be
X'= \frac{\partial}{\partial t} + X
\ee
and its integral lines (called  the orbits of $\omega'$) are precisely (the graphs of the) physical motions $(t,q^i(t),p_i(t))$. Let $\Gamma$ be the space of these orbits and $\pi$ the projection that sends each point of $\cal C$ to the orbit to which it belongs.   $\Gamma$ carries a symplectic two-form $\sigma$ uniquely characterized by the fact that its pull back to $\cal C$ by $\pi$ is $\omega'$. (A pull back is degenerate in the directions of the orbits.)  The symplectic space $(\Gamma,\sigma)$ is clearly the same as the one constructed above. 

The equivalence between the conventional hamiltonian and the presymplectic formulation, is almost complete. The reason for the ``almost" is subtle, interesting, and at the core of the problem discussed in this paper. Given a hamiltonian system $(\Gamma_0, \omega, H)$, we can immediately construct its corresponding presymplectic formulation $({\cal C}, \omega')$.  But the opposite is not true, since we need to know which one of the variables on $\cal C$ is the time variable, in order to do so. In other words, the presymplectic formulation leads to the same relations between the variables $(t,q^i,p_i)$ as the hamiltonian one, but without specifying which of these variables is to be recognized as the time variable. The difference is the same as the difference between giving a function $y(x)$ or its parametrized form $(y(s),x(s))$: in the first case $x$ is singled out as the independent variable, in the second case it is not. 

\subsubsection{General covariant mechanics}

Systems like general relativity, or a single free relativistic particle, are defined in the covariant language by a Lagrangian that leads to a vanishing canonical Hamiltonian.  Equivalently, they are defined by equations of motion that are gauge invariant under a re-parametrization of the evolution coordinate.  The Legendre transform of the Lagrangian of these systems defines a phase space with constraints, and the dynamics is coded in the constraints. Let $\cal E$ denote this phase space (to distinguish it from the phase space of a conventional system, since it has a different physical interpretation) and let $\cal C$ denote the subspace of $\cal E$ where the constraints vanish. $\cal E$ is a symplectic space with symplectic form $\omega$. Its restriction to $\cal C$ is a presymplectic two-form $\omega'$ (the pull back of $\omega$ under the embedding $i$ of $\cal C$ in $\cal E$), which is degenerate in the directions of the hamiltonian vector fields of the constraints themselves.  The space of the orbits $\Gamma$ is again a symplectic space carrying a symplectic two-form $\sigma$, uniquely characterized by 
\be
      i_*\omega = \omega' = \pi_* \sigma. 
\ee 
where
\be
   {\cal E} \stackrel{i}{\longleftarrow}  {\cal C}  \stackrel{\pi}\longrightarrow \Gamma. 
\ee 
The presymplectic constraint surface $(\cal C,\omega')$ defines the dynamics precisely as in the presymplectic formulation of the hamiltonian dynamics described above.  Notice that it defines all the physical correlations among dynamical variables, without specifying one of these as the independent time variable. The distinctive feature of the general covariant systems is therefore to define dynamics as a ``democratic" correlation between variables instead of as evolution with respect to a singled out independent variable.

A simple example is provided by the dynamics of a free relativistic particle. The extended phase space  $\cal E$ is 8-dimensional, with coordinates $(x^\mu,p_\mu)$ and $\omega=dx^\mu\wedge dp_\mu)$. The constraint surface $\cal C$ is given by $p^2=m^2$. The orbits are given by 
\be
      x^\mu(\tau)=\frac{p^\mu}m \tau +  x_o^\mu.
\ee
And there is a six dimensional space of these. Each orbit determines a correlation between observables.  For instance it determines the relation between different coordinates on Minkowski space.  Notice that all this is Lorentz invariant.  Notice also that this canonical formulation never specifies one particular Lorentz time as the preferred one. To obtain a conventional Hamiltonian formulation we have instead to select a Lorentz frame and choose one variable, say $x^0$ (as opposed to $\tilde x^0=\Lambda^0_\mu x^\mu$ where $\Lambda$ is Lorentz matrix) as the time variable.  Then this determines a Hamiltonian $H=\sqrt{\vec p^2+m^2}$, which generates the same motions, but in a non-manifestly Lorentz invariant language.  

Notice that any Gibbs state 
\be
   \rho\sim e^{-\beta H}=e^{-\beta \sqrt{\vec p^2+m^2}}
\ee
breaks Lorentz invariance and selects a preferred Lorentz time. Physically, this is the specific Lorentz-time flow with respect to which a given gas of relativistic particles is in equilibrium.

Notice that it is somewhat misleading to state that the full dynamics of a generally covariant system is entirely captured by the physical phase space $\Gamma$ and all functions on $\Gamma$, because this would be like saying that the dynamics of a harmonic oscillator is captured by writing down the phase space coordinated by $A$ and $\phi$, and all functions of $A$ and $\phi$.  If we do so, we loose track of the fact that the harmonic oscillator is characterized by the oscillating variable $q(t)$! The dynamics of a generally covariant system is not just described by $\Gamma$ and the family of all functions on $\Gamma$.  We also need to give explicitly the embedding of each orbit in $\cal C$ or, equivalently, in $\cal E$. In the case of the relativistic particle, for instance, the dynamics is not just the specification that the physical space is six dimensional: it is also the information that each point of this space determines a timelike line in Minkowski space, namely a correlation between quantities on $\cal C$. In this context, such quantities are called ``partial observables" \cite{Rovelli:2001bz}.

\subsubsection{Statistical mechanics}

The symplectic form defines a volume-form on $\Gamma_0$, obtained by taking $N$ times the wedge product of $\omega$ with itself. This defines an integral on $\Gamma_0$, which we indicate simply without measure notation.  A statistical state is a real non-negative function $\rho$ on $\Gamma_0$ normalized as
\be
\int \rho = 1.
\ee
Its entropy is defined by the Shannon expression
\be
S=-\int \rho \ln\rho.
\ee
The mean value of an observable $A$ in the state $\rho$ is defined by 
\be
\bar A = \int A \rho.
\ee
The mean value of $A(t)$ can be equally obtained as the mean value of $A$ on the state $\rho(t)$ which satisfies 
\be
\frac{d\rho(t)}{dt} = \{\rho,H\}.
\ee
An equilibrium Gibbs state, is a particular statistical state of the form 
\be
 \rho \propto e^{-\beta H}
 \label{gs}
\ee
where $\beta=1/kT$ is a positive real number and $T$ is the temperature. It is immediately clear that a Gibbs state is time independent and the mean value of all observables in a Gibbs state are time independent. Nontrivial time correlations can nevertheless be defined from quantities like 
\be
f_{AB}(t) = \int A(t)B(0) \rho.
\ee
The proportionality factor in \eqref{gs} is determined by the normalization condition: 
\be
 \rho =\frac{1}{Z(\beta)} e^{-\beta H}
\ee
where 
\be
Z(\beta)=\int e^{-\beta H} = e^{-\beta F}
\ee
is called the partition function, and $F$ is called the free energy. It follow immediately from the definitions and a short calculation that the mean value $E$ of the energy is given by 
\be
E=-\frac1\beta\frac{d \ln Z}{d\beta}
\ee
and 
\be
S=\beta E - \beta F. 
\label{thermorel}
\ee
these are the basic thermodynamical relations for the Gibbs states. 

\subsubsection{General Covariant Statistical Mechanics}

Here a condense the results of this paper.  A statistical state is a normalized positive function on the physical state space. It determines a thermal flow with generator $X$ defined by
\be
\rho\ \omega(X)=d\rho.
\ee
the generator of this flow is the (state dependent) thermal hamiltonian $h=-\ln\rho$ and the thermal time $\tau$ is the parameter of this flow. For a conventional Gibbs state in a non-generally-covariant system, temperature is the ratio between thermal time and geometrical time. 

In a gravitational field theory, if $h$ is local, then it defines a flow in spacetime, and a preferred foliation of the mean geometry.  The local temperature, which satisfies the Tolmann relation, is the local ratio between the spacetime flow and proper time.

\subsection{Quantum theory}

\subsubsection{Quantum Mechanics}

A conventional quantum system is defined by a Hilbert space $\cal H$ and a family $\cal A$ of observables $A$, self-adjoint operators on $\cal H$, which in particular includes a Hamiltonian $H$. The Hamiltonian generates a unitary flow on $\cal H$ by the one-parameter group of unitary transformations $U(t)=e^{-iHt}$ in the Schr\"odinger picture and a flow on the observables by $A(t)=U(-t)AU(t)$ in the Heisenberg picture. In the Schr\"odinger picture, the Hilbert space $\cal H$ corresponds to the phase space $\Gamma_0$ at a given time; while in the Heisenberg picture the Hilbert space $\cal H$ corresponds to the phase space $\Gamma$ of the solutions of the equations of motion.  The expectation value of an observable in the state $\psi\in{\cal H}$ is given by $\bar A=\langle \psi A \psi\rangle$, or equivalently by
\be
\bar A=\tr[A\rho]
\label{meanv}
\ee
where 
\be
\rho=|\psi\rangle\langle\psi|.
\label{pure}
\ee
The eigenvalues of $A$ determine the quantization, namely the possible outcomes of a measurement, of $A$ and transition probabilities between such measurement  outcomes are determined by the matrix elements of $U(t)$ in the observable's eigenbasis.

\subsubsection{Quantum Statistical Mechanics}

A statistical state $\rho$ is a trace-class operator on $\cal H$ normalized by 
\be
\tr[\rho]=1.
\ee
The mean value of an observable in such a state is still given by \eqref{meanv}. The states of the form \eqref{pure} satisfy $\rho^2=\rho$, are called ``pure", and their conventional physical interpretation is that the probabilistic nature of the uncertainty in the predictions derived from them is not due to our ignorance, but to irreducible intrinsic quantum uncertainty.  The von Neumann entropy of the state $\rho$ 
\be
S=-\tr[\rho\ln\rho]
\ee
vanishes on pure states. A Gibbs state is a state of the form $\rho\propto e^{-\beta H}$. The partition function is the inverse of its normalization, namely 
\be
Z(\beta)=\tr[-e^{\beta H}].
\ee
Again the basic thermodynamical relation \eqref{thermorel} follows in a few steps from these definitions. 

\subsubsection{General Covariant Quantum Mechanics}

A generally covariant quantum system is defined by an extended Hilbert space $\cal K$, a (possibly generalized\footnote{Namely a subspace of $\overline{\cal S}$ in a Gelfand triple $\overline{\cal S}\supset{\cal K}\supset{\cal S}$.}) subspace $\cal H$, the ``space of solutions of the Wheeler-deWitt equation", and a family of observables $A,B$ on $\cal K$ called ``partial observables".  

The eigenvalues of the partial observables determine the quantization, namely the possible outcomes of a measurement \cite{Rovelli:1992vv,Rovelli:1994ge,Rovelli:2007ep}, and transition probabilities between such measurements' outcomes are determined by the matrix elements 
\be
\langle q |P| q' \rangle
\ee
of the (generalized) projection 
\be
P:{\cal K}\to{\cal H}, 
\ee
in the observables' eigenbases $|q\rangle$ in $\cal K$ (see \cite{Rovelli:2004fk}, Chapter 5 and \cite{Rovelli:2011mf}). A specific  example of a definition of these transition amplitudes, finite to all orders, is provided by covariant loop quantum gravity  \cite{Rovelli:2011eq}. The quantum mechanics of generally covariant systems can therefore be well defined without the need of specifying a time variable. 

\subsubsection{General Covariant Statistical Quantum Mechanics}

The thermal-time flow of a generally covariant statistical quantum state $\rho$ is defined by its Tomita flow. This can be constructed as follows.  The expectation value of a statistical state $\rho$ on the algebra $\cal A$ of the gauge-invariant observables $a$ define a state on this algebra. Assuming $\cal A$  to be a $C^*$-algebra, the GNS construction defines a Hilbert space $\cal H$ where observables are represented by operators and $\psi$ is a vector (even if $\psi$ is a statistical state). Let then $S$ be the operator defined by $Sa\psi=a^*\psi$. It is always possible to write $S$ in the form $S=Je^{h/2}$, where $J$ is antinunitary and $e^{h}$ is self-adjoint. The Tomita flow on the algebra is then defined by 
\be
                 \alpha_t a = e^{-ith} a\;  e^{ith}
\ee
and the Tomita theorem states that this is a one-parameter group of automorphisms of the algebra. 

To understand what is going one, start from a normal quantum field theory. Pure states are vectors in Fock space. Mixed states are density matrices, namely trace class operators $\rho$ on Fock space. These form an Hilbert space, which we can call $\cal H$: notice that a statistical state $\rho$ is now represented by a \emph{vector} in this Hilbert space, for which a convenient notation is $|\rho\rangle$. If $a$ is an observable on Fock space, we can represent it on $\cal H$ as $a|\rho\rangle=|a\rho\rangle$, which is again trace class. If $\rho$ is a Gibbs state for a Hamitonian $H$ at inverse temperature $\beta$, namely $\rho=e^{-\beta H}$, then a straightforward calculation shows that $J| k \rangle= | k^* \rangle$ and $e^{h}|k\rangle=|e^{-\beta H}k e^{-\beta H}\rangle$ satisfy the definition of $S$. Therefore the Tomita flow of the Gibbs state is precisely the time flow scaled by the temperature: $  \alpha_t a = e^{-it(\beta H)} a  e^{it(\beta H)}$.  In other words, the Tomita relation between a state and a flow is the quantum field theoretical version of the classical relation between a state on phase space and its Hamiltonian flow. The operator $J$ flips creation and annihilation operators of the quanta over the thermal state, and therefore codes the split between positive and negative frequencies. For a more detailed discussion, see \cite{Connes:1994hv}. Time flow is fully coded into the statistical state. The local relation between thermal time $d\tau$ proper time $dt$ and temperature $T$ is given by equation \eqref{temperature2}.

\vspace{2mm}

\centerline{---}
\vspace{2mm}

Thanks to Hal Haggard for a careful reading of the manuscript and useful comments.


\end{document}